%% file: artILJS_astro_ph_2015.tex
\documentclass{emulateapj}

\usepackage{amsmath}
\usepackage{natbib}
\usepackage[colorlinks=true,citecolor=blue,urlcolor=magenta,breaklinks]{hyperref}
\usepackage{cool}
\usepackage{eulervm}
\begin{document}
\title[Nearby stars as gravitational wave detectors]{ 
Nearby stars as gravitational wave detectors}
\author{Il\'\i dio Lopes~\altaffilmark{1}, Joseph Silk~\altaffilmark{2,3}}
\altaffiltext{1}{
Centro Multidisciplinar de Astrof\'{\i}sica - CENTRA, Departamento de F\'\i sica, 
Instituto Superior T\'ecnico - IST, Universidade de Lisboa - UL, 
Av. Rovisco Pais 1, 1049-001 Lisboa, Portugal; ilidio.lopes@tecnico.ulisboa.pt}
\altaffiltext{2}{Institut d'Astrophysique de Paris, UMR 7095 CNRS, Universit\'e Pierre et Marie Curie, 
98 bis Boulevard Arago, Paris F-75014, France; silk@astro.ox.ac.uk} 
\altaffiltext{3}{Department of Physics and Astronomy, 3701 San Martin Drive, The Johns Hopkins University, 
Baltimore MD 21218, USA} 
 

\begin{abstract} 
Sun-like stellar oscillations are excited by turbulent convection and have been discovered in some 500 main sequence and sub-giant stars and in more than 12,000 red giant stars.  When such stars are near gravitational wave sources,  low-order quadrupole acoustic modes are also excited above the experimental threshold of detectability, and  they can be observed, in principle,  in the acoustic spectra of these stars. Such stars form a set of natural detectors to search for gravitational waves over a large spectral frequency range, from $10^{-7}$ Hz to $10^{-2}$ Hz. 
In particular, these stars can probe the $10^{-6}$ Hz -- $10^{-4}$ Hz spectral window which cannot be probed by current conventional gravitational wave detectors, such as  SKA and eLISA.  The PLATO stellar seismic mission will achieve photospheric velocity
amplitude accuracy  of $~ {\rm cm/s}$. For a gravitational wave search, we will need to achieve accuracies of the order of $10^{-2}{\rm cm/s}$, i.e., at least one generation beyond PLATO.
However, we have found that multi-body stellar systems have the ideal setup for this type of gravitational wave search. 
This is the case for triple stellar systems formed by a compact binary and an oscillating star. Continuous monitoring of the oscillation spectra of these stars to a  distance of up to a kpc could lead to the discovery of gravitational waves originating in our galaxy or even elsewhere in the universe. Moreover, unlike experimental detectors, this observational network of stars will allow us to study the progression of gravitational waves throughout space. 
\end{abstract}
 

\keywords{asteroseismology --  binaries: close -- binaries:  general -- 	gravitational waves -- stars: general  -- Sun: helioseismology}

  
\section{Introduction}

Astronomers had not expected  that stars, other than the Sun, could provide us with the observed wealth of high quality seismic data, surpassing in diversity and quantity the data gathered from the Sun itself. Previously, the GOLF experiment aboard the 
{\it Solar and Heliospheric Observatory} mission~\cite{1995SoPh..162...61G}  
measured the global acoustic oscillations of the Sun from a spatially unresolved entire solar disk of  Doppler velocity data~\citep{2004ApJ...604..455T}. Presently, following an identical strategy, the COROT and {\it Kepler} space missions~\citep{ 2006ESASP1306...33,2010ApJ...713L.160G}, by measuring the light integrated from the entire visible star's surface,  have searched for global oscillations in more than 150,000 main sequence, subgiant and red giant stars~\citep{2009IAUS..253..289B,2011ApJ...738L..28V}. 
A sample of more than five hundred low mass subgiants and main sequence stars were discovered to have  rich Sun-like acoustic oscillation spectra filled with tens of velocity amplitude peaks~\citep{2014ApJS..210....1C}. 
Similar oscillations have been found in more than 12,500 K-G giant stars. 
Although their acoustic spectra are very distinct from those of  main sequence stars, 
as a consequence of their quite distinct structure~\citep{2011MNRAS.414.2594H}, 
these oscillations still qualify as Sun-like oscillations.
As in the Sun, these oscillations are excited by turbulent convection  
and  intrinsically damped by the radiation and the turbulence 
of the outer layers of the star. 
Such seismic observational surveys are able to follow in a systematic and continuous manner the pulsation spectra of the Sun and stars in the solar neighbourhood within a range of up to
one thousand parsecs distance.
These surveys observed fields of stars in the galactic plane, as well as in a few directions above and below it~\citep{2012EPJWC..1905012M}, 
as in the Kepler mission which has been observing stars 13$^o$ above the galactic plane in
the Cygnus region. 
This local gravitational system of celestial bodies made of many stars of different sizes and masses, constituted  of single stars,  binaries or multi-stellar systems, as well as a few stellar clusters~\citep{2012ApJ...757..190C}, is a natural network of detectors for gravitational radiation.
In this work we argue that this natural network of stars will allow us to search for gravitational wave imprints in the oscillation spectra of single stars, or for contemporaneous signatures of the same gravitational wave event on the spectra of two or more close stars.
It will also allow us to follow the progression of a gravitational wave through space, 
across the spectra of several stars - when it passes through a stellar cluster -  
and to monitor its progression as it approaches the Earth.

\smallskip
The most likely source of gravitational waves able to stimulate quadrupole modes in Sun-like stars are short-period binaries of two compact objects such as white dwarfs, or even neutron stars and massive black holes.  Low mass binaries are the leading gravitational source candidates, despite having relatively weak gravitational wave emission, as they are numerous and are located in close proximity to the Solar System~\citep{2013GWN.....6....4A}. 
More than 50 ultra-compact binaries, with a size of about a fraction of the solar radius and a period shorter than one hour, have been discovered at a distance between 50 and 700 parsec, an example 
being the cataclysmic variable star AM CVn located at a distance of 606 parsecs from Earth~\citep{2007ApJ...666.1174R}. AM CVn is a binary system where a white dwarf accretes matter from a companion star, leading to the formation of an accretion disc with continuous strong emission in UV and X-rays, and with occasional outbursts. 
If seismic surveys are set to observe and monitor stars  
near these binaries, this could lead to the discovery of gravitational waves. 

The idea that gravitational radiation could excite the normal modes of vibration of  celestial bodies such as the Earth and the Sun was originally discussed by Dyson~\citep{1969ApJ...156..529D} and many other papers  have  followed up this idea. 
Most recently~\citet{2014MNRAS.445L..74M} have estimated the gravitational radiation that is absorbed by stars and~\citet[][]{2011ApJ...729..137S} are among others to suggest the impact of gravitational waves on solar oscillations. Gravitational wave detection through stars and resonant mass detectors~\citep{2006CQGra..23S.239A,2007PhRvD..75b2002G} both work based in a similar principle. 
In the latter case, the detection is done by accurately measuring the tiny variation of the detector size
when one or more modes are excited by a passing gravitational wave~\citep{2009LRR....12....2S}. 
In stars detection is feasible  by monitoring the variations of velocity of the modes at the surface. 
Some aspects of the analysis for stars are similar to the case of spherical resonant mass detectors.

\section{The acoustic spectra of the Sun and stars}
 
Stars, like musical instruments, vibrate in a multitude of eigenmodes. The discrete sequence of frequencies for stellar oscillations can be labelled by two independent integers: the degree of the mode $l, $ which is the degree of the spherical harmonic related with the horizontal eigenfunction; and the order of the mode $n$ which measures the number of nodes of the radial eigenfunction along the radius. For each value $l,$  there is a sequence of resonant acoustic modes that are labelled with $n.$ 
The latter ones correspond to the fundamental tone and overtones of a musical instrument. 
Whole-disk observations of stars through the seismic space missions detect only the large luminosity variations in the stellar surface; therefore, these observations are only sensitive to the lowest values of $l$ ($l\le 3$). The angular frequency of such low-degree acoustic modes satisfies  
\begin{eqnarray}
\omega_N=(n+1/2\; l+\alpha_o)\bar{\omega}+ \zeta_N     
\label{eq:nuN} 
\end{eqnarray} 
where $N$ is a subscript that defines a specific eigenmode $N\equiv (n,l)$, $\alpha_o$ and $\bar{\omega}$ 
are constants and  $\zeta_N $ is a second order term that can be neglected  when $n$ is large~\citep{2001MNRAS.321..615L}.
The constant $\bar{\omega}$ is relates do $\nu_o$ by $\nu_o=\bar{\omega}/2\pi$. This last quantity is known as the large separation.
The frequencies with the same degree are separated apart by $\nu_o$. This quantity is equal to $1/\left(2\int_0^R dr/v_s(r)\right)$, the time taken for the sound wave to travel with a speed $v_s (r)$  from the surface to the center of the star and return.  According to  equation (\ref{eq:nuN}) for $\omega_{N}$,  $\nu_o$ is equal to $(\omega_{(n,l)}-\omega_{(n-1,l)})/2\pi$.
The previous equation with minor adjustments 
has been shown to be valid for many main sequence and red giant stars~\citep{2011A&A...525L...9M,2012ApJ...757..190C}.
Table~\ref{tab:nusun} lists the frequencies of the quadrupole mode as measured by the GOLF experiment. 
When no observational data is available, we show in italics the predicted values 
for the standard solar  model~\citep{2013ApJ...765...14L}. 
One can notice that for frequencies of the acoustic modes of low degree,
the disagreement between 
theory and observation is at most of the order of a 
few percent~\citep[see ][and references therein]{2012RAA....12.1107T}.
Table~\ref{tab:nusun} lists the frequencies for the quadrupole modes.
  
Recent observations have shown that many main-sequence, subgiant and red giant stars 
have a spectrum of acoustic oscillations identical to the Sun, with only minor differences~\citep{2011A&A...525L...9M,2014ApJS..210....1C}.      
The properties of a star's spectrum are characterized by three main quantities: 
the large separation, the frequency at which the amplitude of the spectrum of oscillations is maximum,
and the effective temperature of the star. 
At present, when these quantities are available,
they provide the most reliable method for determining the mass and radius of a star~\citep{2013ARA&A..51..353C}. 
It has been shown from observational data, that homology ratios for  mass, radius and effective temperature hold between the Sun and these stars.  
In particular, the large separation scales as:
\begin{eqnarray}\left(\frac{\bar{\omega}}{\bar{\omega}_\odot}\right)= \left(\frac{M}{M_\odot}\right)^{1/2} \left(\frac{R}{R_\odot}\right)^{-3/2}, \label{eq:baromega}\end{eqnarray}   
where $M$ and $R$  are the mass and radius  of the star.
Equally, $\bar{\omega}_\odot/2\pi$, $M_\odot$ and $R_\odot$ are the solar equivalent quantities.  
In the case of the Sun, the mean large separation $\bar{\omega}/2\pi$ 
is of the order of $135 \;\mu$ Hz (as computed from table \ref{tab:nusun}).
 
\input{table_nusun_article}
\section{Interaction between gravitational waves and acoustic modes}

Within the  framework of general relativity, far from the source of gravitational radiation, the space-time metric tensor  is distorted relative to  flat spacetime (Minkowski) value by a very small spatial component, $h_{\rm ij}$.  In a Galilean coordinate frame whose origin coincides with the center of the star, the stellar material experiences a force proportional to $h_{\rm ij}$. 
Thus, the quadrupole modes of vibration of the star will be excited by gravitational waves when the frequency of the incoming  waves is close to the eigenfrequency of the modes. 
In this case, the time variation of the amplitude of the mode of vibration is described by a harmonic damped oscillator with an excitation source proportional to $h_{\rm ij}$.  As usual, we express the tensor $h_{\rm ij}$ as the sum of spherical components  $h_m$, for which the $m$ (azimuthal order) is an integer such that $|m|\le l$.    The dynamics of general relativity implies that only non-radial modes of degree larger that two can be excited. Of these, the forcing of the quadrupole modes is normally the greatest. Actually, the differential rotation in the Sun and Sun-like stars can also split the $\omega_N$ (equation \ref{eq:nuN}) leading to a subset of $2m+1$ frequencies for each $l,$   which in the case of  quadrupole modes corresponds to five values differing between them only by a few  microHz. Nevertheless, as we are only concerned about the amplitude of the modes, in this study the solution will be found for a generic value of $m.$  
As we restrict our attention to quadrupole acoustic modes of order $n$ with fixed (undistinguished) 
$m$ for which  $|m|\le 2$, in the remainder of this article the fiducial mode will be represented
as $N$ or simply by $n$, meaning $N\equiv (n,2,"m")$.

The strength by which the quadrupole modes of the star are stimulated by gravitational radiation 
depends on the absorption cross-section for gravitational radiation
or its integrated value (in frequency), the quality factor,
and the amplitude of the root mean square velocity at the surface of the star, 
also known as the photospheric velocity.
The theoretical calculation of these quantities is computed in an identical manner
to  gravitational wave detectors.  
Accordingly, the impact of a plus polarized monochromatic gravitational wave,
such as $h_\star\cos{(\omega t)}$, with a strain $h_\star$ and frequency $\omega$, 
averaged over several cycles is estimated as follows:  

- First, the absorption cross-section for gravitational waves by the star, 
$\sigma_{\rm abs}(\omega)$, is defined by expressing the balance 
between the amount of energy $E_{\rm abs }$ which is absorbed by the star 
 and the amount of incident energy  $E_{\rm in}$ on the star's surface:
\begin{eqnarray}  
\frac{dE_{\rm abs}}{dt}=\sigma_{\rm abs}(\omega) \frac{dE_{\rm in}}{dAdt}.
\label{eq:Eabs}
\end{eqnarray}     
where $dE_{\rm in}/dAdt$ is the energy arriving per unit of time, per unit  area.

- Second, the average $dE_{\rm in}/dAdt$ value  for the case of a monochromatic gravitational wave 
is equal to $c^3h_\star\omega^2/(32\pi G)$ where c and G are the values of the velocity 
of light in vacuum and Newton's gravitation constant. Moreover, the gravitational wave averaged over several cycles
leads to $\langle \cos{(\omega t)}\rangle$, which is equal to 1/2.     

- Third, the energy $E_{\rm abs }$ that is absorbed by the star per unit of time by each resonant mode $N$ averaged for a few cycles, is equal to the product of the gravitational wave force $F_{\rm gw} (t)$ and  the velocity of the mode $\dot{\xi}_N$ where $\xi_N$ is the amplitude of the stimulated quadrupole mode.  
The gravitational force $F_{\rm gw} (t)$ is equal to  $M_N L_n \ddot{h}_m $,
where $M_N$ and $L_n$ are the  modal mass  and  the modal length of the mode,  both of which depend on the properties of the acoustic eigenfunction,  
and $\ddot{h}_m$ is the second derivative of $m$'s spherical component, the perturbation tensor $h_{ij}$.
The 
modal length of the mode $L_n$ is an effective length of the mode equivalent to the size of a resonant detector. $L_n$ is unique for each acoustic mode, $L_{n}=1/2\; R\; |\chi_{n}|$ where    $\chi_{n}$ is a coefficient that depends of the eigenfunction $\xi_N$:   
\begin{eqnarray} 
\chi_{n}=\frac{3}{4\pi \bar{\rho}_{\star}} \int_0^1 \rho (r)\left[ \xi_{r,n}(r)+3 \xi_{h,n}(r) \right] r^3 dr. \label{eq:chin} 
\end{eqnarray}   where $\rho$ and $\bar{\rho}_{\star}$ are the density profile and mean density of the star,  
and $\xi_{r,n}$ and  $\xi_{h,n}$ are the radial and horizontal components of the quadrupole modal eigenfunction $\xi_{N}$, respectively.  
As originally computed for a resonant detector~\citep[e.g.,][]{Maggiore:2008tk},  
the absorbed energy per unit time is expressed as  
\begin{eqnarray} \frac{dE_{abs}}{dt} \equiv \langle F_{\rm gw}(t) \dot{\xi}_N  \rangle = \frac{M_N h_\star^2 \;L_{n}^2\; \eta_n \omega^6}{(\omega^2-\omega_{n}^2)^2+4\eta_{n}^2\omega^2},  
\label{eq:Eabshm} 
\end{eqnarray}  where $\eta_n$ is the damping rate of the mode.    

\smallskip

Equations (\ref{eq:Eabs}) and (\ref{eq:Eabshm}) relate to the energy of the incident gravitational wave with the energy absorbed by the acoustic mode $N$, from which is obtained   the average absorption cross-section:\begin{eqnarray}  \sigma_{\rm abs}(\omega) =  \frac{M_N  \eta_n L_{n}^2\;\omega^4}{(\omega^2-\omega_{n}^2)^2+4\eta_{n}^2\omega^2}\; \frac{32\pi G}{c^3}  \label{eq:sigmaN} \end{eqnarray}   

The response of a mode to a gravitational perturbation is better evaluated by the integral of the absorption cross-section, $\Sigma_n=\int\sigma_{\rm abs}(\omega) d\omega/2\pi$ for which the limits of the integral correspond to the frequency interval of the gravitational wave-packet. However, as the function $\sigma_{\rm abs}(\omega) $ only is significantly different from zero near each resonance frequency  $\omega_n$, conveniently, without much loss of accuracy, the limits of the integral can be replaced by $-\infty$ and $+\infty$.  For the same reason, the second term in  the right-side of the equation (\ref{eq:sigmaN}) is expanded in $(\eta_n/\omega_n)$.

$\Sigma_n$  near the resonance frequency,  reduces to 
\begin{eqnarray}  \Sigma_n=\pi\chi_{n}^2 \; \frac{M_N G}{c}\; \left(\frac{R\omega_n}{c}\right)^2, 
\label{eq:sigmaNf} \end{eqnarray}
thus, $M_N=E_N M$ where $E_N$ is the normalized inertia of the mode.
This approximate result shows that the integrated cross-section is independent of the quality factor of the mode, $Q_n\equiv\omega_n/(2\eta_n)$. This occurs because at the peak $\sigma (\omega)$ is proportional to  $Q_n$, so that  $Q_n$ cancels in $\int d\omega \sigma (\omega)$. Using equation (\ref{eq:nuN}),      
$ \Sigma_n=F_n\;({M_N G}/{c}) ({R\bar{\omega}}/{c})^2 $ 
where $F_n=\pi\chi_{n}^2 \; (n+1+\alpha_o)^2$. In the particular case that the sound speed 
is constant, for instance equal to the average sound speed $\bar{v}_s$,  $\bar{\omega}=\pi\bar{v}_s/R$.   
$\Sigma_n$ becomes identical to the one computed for a resonant mass detector, 
$ \Sigma_n=F_n^{\star}\;({M_N G}/{c})\; ({v}_s/{c})^2$ where $F_n^{\star}=\pi^3\chi_{n}^2\;(n+1+\alpha_o)^2$~\citep{Maggiore:2008tk}.

\begin{figure*}
\centering 
\includegraphics[scale=0.8]{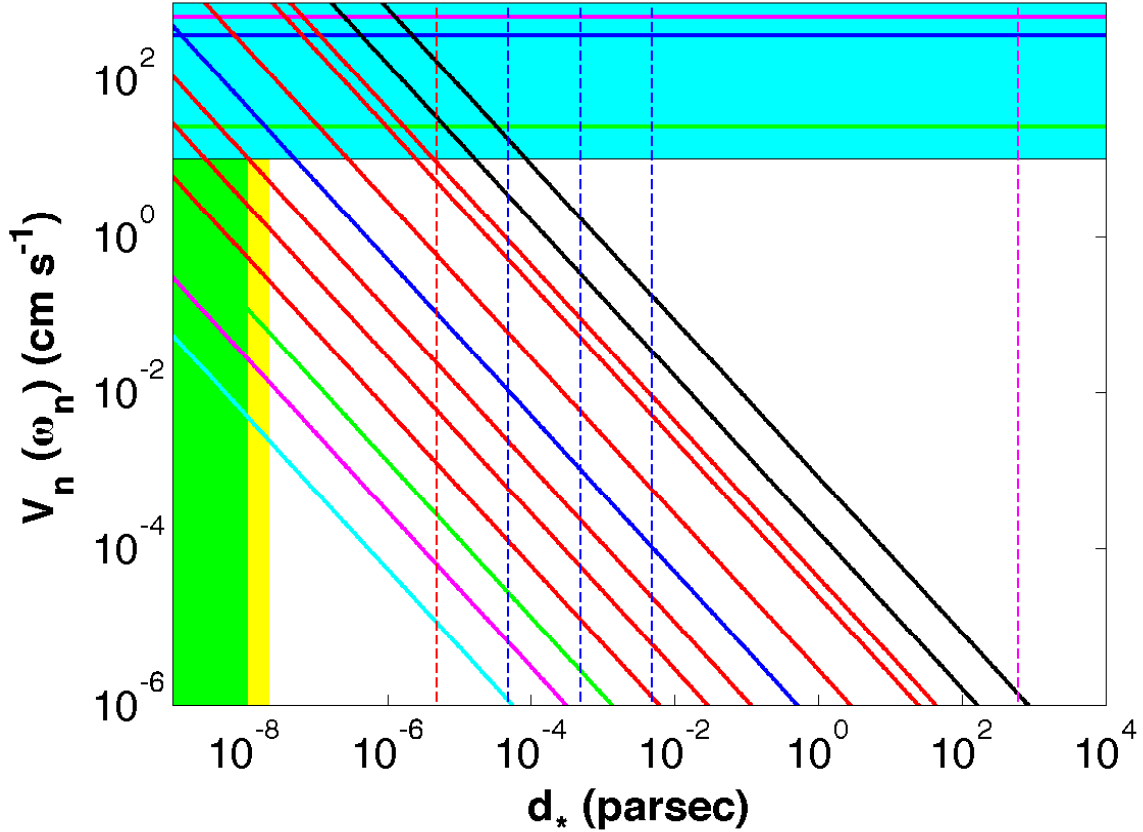}
\caption{ 
Impact of a gravitational wave in a {\it Sun-like star detector}: 
The  photospheric velocity  $V_{n} (\omega_n)$ (in $cm\;s^{-1}$) for quadrupole acoustic modes 
is a function of the distance to the gravitational wave source,  as predicted by equation (\ref{eq:VNgwf}).
In the case of modes with $n\le 5$ (see Table~\ref{tab:nusun}), the photospheric velocity
is computed in the undamped limit, i.e., the photospheric velocity corresponds to  $V_{n,u}={T^\star_n}^{1/2}\;V_{n}$. 
As an example of a gravitational wave source,
we choose the compact binary system AM CVn~\citep[e.g.,][]{1998A&A...332..939S,2007ApJ...666.1174R}
located at a distance $d_{\earth}=606$ pc, which has a size smaller than $0.5$ of the solar radius 
and a orbital period of $\sim 1029 s$, 
for which the strain amplitude for Earth detectors is predicted to be $h_{\earth}=2\;10^{-22}$.
Accordingly, the strain amplitude $h_\star$ of the arriving gravitational wave to the Sun-like star 
is computed as $h_\star=3 \;10^{-14}/d_{\star}(AU)$. 
The vertical dashed-lines correspond to the distance of the gravitational
wave source to the {\it star's detector}: Sun (magenta) and a Sun-like star  
located at a distance of 1 AU (red) or 10, 100 and 1000 AU (blue).
The vertical green and yellow areas correspond to $0.5$ and $1.0$ of the solar radius.  
The oblique lines correspond to the quadrupole modes: 
$p_i$ with $i=0,1,2,4,5,6$ (red),  $p_i$ with $i=3,7,12,18$ (blue, green, magenta, cyan)
correspond to  $h_\star=3 \;10^{-12}/d_{\star}(AU)$ with $h_\earth=10^{-20}$ (see Table~\ref{tab:nusun}).
For reference, we show $p_0$ and $p_1$ (black lines) corresponding to  
the steady-state (or saturated) limit. 
The horizontal cyan band ($V_{n}\ge 1.0\; cm\;s^{-1} $) defines the expected $V_{n}$ threshold  
of the Plato mission. To illustrate the variety of sun-like star oscillations, 
we show several maximum $V_{n} $ measurements (horizontal lines): 
Sun (green), Procyon (blue) and $\nu$ Ind (magenta) stars (see main text). 
See Table~\ref{tab:nusun} for further information about quadrupole modes.}
\label{fig:A}
\end{figure*}

The relevant observable that will allow us to identify the impact of gravitational waves
in an oscillation mode is the   
phostopheric velocity that is computed as $V_{n}^2 (\omega_n)\equiv 1/({2\eta_n M_N}) \;dE_{abs}/dt$~\citep{1977ApJ...211..934G}. Therefore, using equation (\ref{eq:Eabshm}), we obtain
\begin{eqnarray} 
V_{n}(\omega_n)
=\frac{h_\star L_n\omega_n^2}{\alpha_s \eta_n}.
\label{eq:VNgwf} 
\end{eqnarray} 
In the previous equation, we introduce the numerical constant $\alpha_s$ 
which is equal to $2\sqrt{2}$, for which the exact value is fixed by observations.   
Since the amplitude of the strain decreases with the inverse of the distance,   
$h_\star$ is computed as $h_\star=d_{\earth}/d_\star\,h_{\earth}$, 
where $d_{\earth}$ and $d_{\star}$ are the distances of the source of gravitational 
radiation to the Earth and to the Sun-like star, and
$h_{\earth}$ is the current strain prediction for the Earth detectors. 
An illustrative example is shown in Figure~\ref{fig:A}.

The photospheric velocity, as shown in equation (\ref{eq:VNgwf}), relates to the excitation of quadrupole modes
of frequency $\omega_n$ by a monochromatic gravitational wave with the same frequency $\omega$, i.e., $\omega=\omega_n$. 
Nevertheless, in some cases as noticed by~\citet{2014MNRAS.445L..74M}, $\omega$ drifts across the frequency $\omega_n$ 
like it occurs during the inspiral phase of binary systems. The magnitude of the frequency drift $\dot{\omega}$ 
is mostly dependent of the chip mass of the binary $M_c$, as $\dot{\omega}\approx 3 \left(GM_c/c^3\right)^{5/3}\omega^{11/3}$~\citep{Maggiore:2008tk}. 
The contribution of $\dot{\omega}$ for the excitation of a quadrupole mode 
is only relevant if the duration of the gravitational time  $\tau_f=1/\sqrt{\dot{\omega}}$
is smaller than the damping time of the mode $\tau_n=1/(2\eta_n)$, or if the ratio 
$T^\star_n\equiv\tau_f/\tau_n$ is smaller than one. 

The steady-state (or saturated) limit corresponds to $T^\star_n \gg 1 $ 
(or $\tau_f\gg \tau_n$) for which  $\dot{\omega}$ can be neglected in the calculation 
of the photospheric velocity (equation \ref{eq:VNgwf}). However, as shown by~\citet{2014MNRAS.445L..74M}
in the reverse case of undamped oscillations, for which  $T^\star_n \ll 1 $  (or $\tau_f\ll \tau_n$),
the slow frequency variation of the gravitational wave must be taken into account,
accordingly $\omega_n$ must be replaced  by $\omega_n+\dot{\omega}_n\tau$,  where $\tau$ is the time difference to the resonance. The calculation of $\tau_f$  is given by 
$4.81\times 10^{9} \left(M_c/M_\odot \right)^{-5/6} \left(\omega_n/{\rm m Hz} \right)^{-11/6}$~\citep{Maggiore:2008tk}.

\citet{2005MNRAS.357..834R} have estimated that the averaged energy transfer 
to quadrupole mode by a slow varying frequency gravitational source is given by $\pi h_\star^2/(4\dot{\omega}_n)$. 
Accordingly, the energy absorbed by a stellar quadrupole mode per unit time  in the undamped limit ($T^\star_n \ll 1 $, subscript $u$) 
is expressed as  $ \left({dE_{abs}}/{dt}\right)_u=T^\star_n\;\left({dE_{abs}}/{dt}\right)$ where
$\left({dE_{abs}}/{dt}\right)$ corresponds to the solution given by equation~(\ref{eq:Eabshm}).
$\left({dE_{abs}}/{dt}\right)_u$ relates with the work done by an external slow varying  
gravitational source (see~\citet{2014MNRAS.445L..74M} for details).
Accordingly, the photospheric velocity in the undamped limit is reduced in relation to the steady-state limit
(i.e., $V_{n}$), it reads  $V_{n,u}(\omega_n) ={T^\star_n}^{1/2}\;V_{n}(\omega_n)$.
Equally, $\sigma_{\rm abs,u}(\omega)$  and  $\Sigma_{n,u}$ are also reduced in the case of the undamped limit, 
as these quantities are obtained by multiplying $\sigma_{\rm abs}(\omega)$ (equation~\ref{eq:sigmaN}) and  $\Sigma_n$  (equation~\ref{eq:sigmaNf}) by the same $T^\star_n$ factor.

The integrated absorption-cross section (equation \ref{eq:sigmaNf}) and the photospheric velocity 
(equation \ref{eq:VNgwf}) for a star can be scaled to the Sun. Using the eigenfrequency equation (\ref{eq:nuN}) 
retaining only the first term, and the scale relation  (\ref{eq:baromega}), 
the integrated absorption-cross section reads
\begin{eqnarray}  
\frac{\Sigma_n}{\Sigma_{n\odot}} 
=  \left(\frac{\chi_n}{\chi_{n\odot}}\right)^{2}
\left(\frac{E_n}{E_{n\odot}}\right)
\left(\frac{M}{M_\odot}\right)^{2}
\left(\frac{R}{R_\odot}\right)^{-1}
\label{eq:Sigmastar}
\end{eqnarray}
and the photospheric velocity reads
\begin{eqnarray}  
\frac{V_{n}}{V_{n\odot}}
=  \left(\frac{\chi_n}{\chi_{n\odot}}\right)
\left(\frac{\eta_{n}}{\eta_{n\odot}}\right)^{-1}
 \left(\frac{M}{M_\odot}\right)
\left(\frac{R}{R_\odot}\right)^{-2}
\label{eq:Vstar}
\end{eqnarray}
where the subscript $\odot$ denotes the Sun. 
The equivalent expressions for the undamped/unsaturated case ($T^\star_n \ll 1 $) are obtained by
multiplying the right-side of both previous equations by the ratio
$(T^\star_n/T^\odot_n)$ and $(T^\star_n/T^\odot_n)^{1/2}$, respectively.

\section{The sensitivity of star's detectors to gravitational waves}

Similarly to convectional resonant mass detectors, the stimulation of quadrupole oscillations in stars by an external gravitational radiation source depends on the integrated cross-section and the quality factor. Table~\ref{tab:nusun} lists these quantities, as well as the amplitude of the photospheric velocity for the quadrupole acoustic modes in the Sun.
Moreover, we also computed the photospheric velocity in the case of undamped oscillations and the ratio $T^\star_n$.
In this study we chose as fiducial gravitational wave source the AM CVn binary system ($M_c=0.248\;M_\odot$),
for which $\tau_f$ is equal to $\sim 1.54\times 10^{10} \; \left(\omega_n/{\rm m Hz} \right)^{-11/6} {\rm sec}$.
Existing observations and other theoretical quantities computed from an updated standard solar model~\citep[e.g.,][]{2013ApJ...765...14L} are also summarized in the table. 
The quadrupole modes that have the largest integrated cross-sections and quality factors, as well as the largest photospheric velocity, are the modes of lower order.
In the following, we give the reasons why Sun-like stars 
are good gravitational wave detectors:

\smallskip
 
- It follows from equation (\ref{eq:sigmaNf}) that the integrated cross-section of a mode depends on 
the modal mass and the length of the mode, as well as the sound speed in the interior of the star. Because stars have much larger masses than convectional resonant mass detectors, their integrated cross-section is many orders of magnitude larger. In the Sun, low order modes have a $\Sigma_n$ that is 17 or 20 orders of magnitude (if we consider the steady-state or undamped limits respectively) larger than the current detectors,
such as Mario Schenberg~\citep{2006CQGra..23S.239A},
Minigrial~\citep{2007PhRvD..75b2002G} 
and Allegro~\citep{1996PhRvD..54.1264M}. 
\citet{2007PhRvD..75b2002G} estimates a $\Sigma_n$
of the order of  $9.8\, 10^{-22}{\rm cm^{-2} \, Hz} $ for a high performance resonant mass detector. 
Even for high order modes ($n\sim 15$) the Sun, as a detector, performs better by a ten-magnitude factor. More massive stars have a fractionally smaller $\Sigma_n$ than the Sun.
Hence, from equation (\ref{eq:Sigmastar}), it follows that $\Sigma_n$ increases with $M^2$ and decreases with $R$, as for most of the observed Sun-like stars, like red giants~\citep{2012A&A...537A..30M}, the smaller mass corresponds to $0.5\,M_\odot$ and the largest radius to $15\,R_\odot$, which leads to a maximum reduction of  $\Sigma_n$ by a negligible  factor of $0.02$.
The variation of $\chi_n$ and $E_{n}$ among these stars also introduces important corrections to $\Sigma_n$,
in many cases, this should increase the value of $\Sigma_n$. Therefore the net result found for the Sun also holds for
many of  these stars.

\smallskip

- Stars respond to gravitational radiation as a truly high Q oscillator. As shown in table~\ref{tab:nusun}, this is particularly relevant in the case of low order modes. Actually, the $Q_n$ for small $n$ is 1 to 2 orders of magnitude better than for current detectors. 
\citet{2007PhRvD..75b2002G} estimates that $Q_n \sim 10^7$ for a CuAl alloy spherical detector. In Sun-like stars, such high Q is due to the predicted low  rates, believed to be caused by the damping processes related with radiation and turbulence convection of the upper layers of the star.
This behaviour is most relevant for main sequence and subgiant stars
(with $M$ varying from $0.95\;M_\odot$ to $2.0\;M_\odot$) for which current theoretical models predict that the $\eta_n$ for low $n$ modes is of the order of $10^{-5}$ to $10^{-7}$~\citep{1999AaA...351..582H, 2005AaA...433..349B}. 
It was also found that for global acoustic modes ($l\le 3$) , that $\eta_n$ is independent 
of $l$ and always increases with $n.$  Besides, $\eta_n$ always decreases with the mass of the star 
and also as its age increases.
This behaviour is similar for red giant stars. Recent seismic data has shown that the average damping rate $\langle \eta_n \rangle$  of Sun-like stars increases with the effective temperature. This relation holds not only for main sequence and 
subgiant field stars, but equally for red giant stars, some of which were discovered in open stellar clusters~\citep[e.g.,][]{2012ApJ...757..190C}.
As  the $\langle \eta_n \rangle$ of red giant stars is two orders of  magnitude smaller than for the Sun's case, we expect that the $\eta_n$ of low order modes varies by a similar amount, leading to an increase of $Q_n$.
Nonetheless, a smaller quality factor could be beneficial, as a lower $Q$ factor increases the possibility of  quadrupole modes being easily excited by gravitational radiation.

\smallskip

- According to equation (\ref{eq:VNgwf}), the  low order quadrupole modes are the ones for which the
photospheric velocity is predicted to have the largest value. 
The velocities of the modes of higher order are not so high. The increase of the photospheric velocity depends of the quantities on the ratio  $\omega_n^2/\eta_n$ which for low $n$ have not yet been measured for the Sun, and only theoretical predictions are available, as shown in italics in table~\ref{tab:nusun}. Nevertheless, both quantities are predicted by theoretical models, which successfully reproduce the observational data at higher frequencies. 
The $L_n$ and its related quantity $\chi_n$ are the other critical quantities 
that affect the photospheric  velocity. 
Unlike for a typical detector, $L_n$ and $\chi_n$ decrease rapidly with $n$. As shown in equation (\ref{eq:chin}), this is due to the fact that the density in the star  decreases with the increase of the radius, 
and the depth reached by the acoustic 
eigenfunction $\xi_N$ decreases as n increases. As shown in table~\ref{tab:nusun},  
$L_{n}$ is only significant for  modes of low order. The decrease of $L_n$ and $\eta_n^{-1}$ with $n$ is the main reason for the rapid decrease of photospheric velocity with $n$. 
Nonetheless, we notice that in the case of the frequency of the gravitational wave drifts during the excitation of the quadrupole mode, the maximum $V_n$ attained will
be reduced by up to several orders of magnitude, as we discuss in the previous section. Table~\ref{tab:nusun} shows the photospheric velocity computed for a strain of $10^{-20}$,  which for the first low order modes, is predicted  to have a maximum amplitude of $10^{-6}\;{\rm cm s^{-1}}$ or $7\;10^{-8}\;{\rm cm s^{-1}}$ respectively for the steady-state ($T^\star_n \gg 1$) and undamped ($T^\star_n\ll 1$) limit cases. In this study, the reduction of $V_{n,u}$ over steady-state
is as much as two orders of magnitude, and will be greatest for modes with the largest cross-sections.  
Indeed, the excitation of quadrupole modes by gravitational waves is limited at low order by the fact that oscillations are unsaturated, and at higher order modes because these have small cross sections 
to the impact of gravitational waves (see Table~\ref{tab:nusun}).

\smallskip

Thus in order for  gravitational waves in the star acoustic spectrum
to be detected, it would be necessary for  $V_n$ to be larger than the current signal-yo-noise ratio instrumental threshold.
A current estimation of this threshold for a future  interplanetary space mission~\citep{2009ExA....23..491A}, made from an average of the 50 modes observed by 
the GOLF experiment~\citep{2004ApJ...604..455T,2007Sci...316.1591G,2009ApJS..184..288J} during a period of 
10 years fixes the observation limit of 1 $\sigma$ level between $10^{-2}{\rm cm\;s^{-1}}$ and $3\,10^{-4}{\rm cm\;s^{-1}}$.
Nevertheless, for the next planed seismic missions, TESS~\citep{2014SPIE.9143E..20R} and Plato~\citep{2013EGUGA..15.2581R} the
 minimum $V_n$ measured is  expected to be of the order of $1\; {\rm cm\;s^{-1}}$.
Although, this precision is not sufficient to detect gravitational waves with current sun-like stars, 
it is possible that goal to be achieved in a next generation of seismic instruments,  
as we show in figure~\ref{fig:A} if a precision of 
$10^{-2}\;{\rm cm\;s^{-1}}$ is attained, this type of detection could be successful in some
specific multi-body stellar systems. In the remaining of the article, this value is used as a detectability threshold for a future space mission.  
Indeed, it follows that the $V_{n}$  for the $f$-mode, $p_1$-mode and $p_2$-mode increases above the threshold of $10^{-2}\;{\rm cm\;s^{-1}}$, if  $d_\star$ is smaller than 10 AU, as illustrated schematically in Figure~\ref{fig:A} 
for the case of a fiducial compact binary as the AM CVn binary. Obviously, if the strain is 2 orders of magnitude larger, it will be much easier to successfully detect the gravitational radiation (cf. Figure~\ref{fig:A}).

The variation of mass and radius of the star affects the amplitude of the photospheric velocity, and  it follows from equation (\ref{eq:Vstar})  that  $V_n$ increases with $M$ and decreases with  $R^2$. 
Thus ignoring the variation of $\chi_n$ and $\eta_n$, and taking the minimum $M$ and maximum $R$ values 
from the recent list of observed Sun-like stars~\citep[e.g.,][]{2012A&A...537A..30M,2014ApJS..210....1C}, at $0.5\,M_\odot$ and $15\,R_\odot$, the photospheric velocity  is reduced by a factor $2\,10^{-3}$. One such example is the star KIC 5822889 for which $V_n$ is $0.01 V_{n\odot}$  For other stars the effect is the reverse, as for
KIC 7970740 for which $V_n \approx 1.3 V_{n\odot}$~\citep{2014ApJS..210....1C}.
This is a rough estimate, because these calculations do not account for the potential differences
coming from $T^\star_n$, $\eta_n$ and $\chi_n$. 
In particular, notice the case of
quadrupole modes with very low $n$ for which the excitation by gravitational waves occurs in the unsaturated case,
$T^\star_n\equiv \tau_f/\tau_n$ reduces further the value of $V_n$. In the case of the AM CVn compact binary, 
$T^\star \sim 10^{-3}$ assuming that $\eta_n$ is similar to the Sun, 
$V_n$ reduces by a further $\sqrt{10^{-3}}$ factor. 
For instance,  $\eta_n$ for low $n$ is 
of the order of $10^{-6}\mu Hz$ and does not change much among this type of star, as $\chi_n$ increases for more massive stars, due to the increase of the density in the star's core. Although Sun-like stars can have different $\chi_{n}$ and $\eta_n$, the low $n$ modes should be always the ones 
with the largest velocity amplitudes. 

Actually, in many spectra of Sun-like stars, acoustic modes were found with large photospheric velocities, 
believed to be stimulated by the turbulent convection in the envelopes of these stars.   
Figure~\ref{fig:A} shows the maximum photospheric velocities 
measured for a few of the most well-known stars~\footnote{Characteristics of a few stars:
Procyon has $M=1.5M_\odot$ and $R=2.0R_\odot$; and $\nu$ Ind has $M=0.85M_\odot$  and   $R=2.9R_\odot$} 
other than the Sun - with $20$ cm$s^{-1}$~\citep{2009ApJS..184..288J}, 
Procyon with $38$ cm$s^{-1}$~\citep[][]{2004A&A...413..251K,2007AaA...464.1059L,2008ApJ...687.1180A,2010ApJ...713..935B} 
and the sub-giant star $\nu$ Ind with 
$650$ cm $s^{-1}$~\citep{1996MNRAS.280.1155B,2006ApJ...647..558B,2007A&A...470.1059C,2008ApJ...682.1370K}. 
This behavior follows the well-known scaling equation for the maximum velocity amplitude of acoustic modes $V_{\rm n,max}$ excited by turbulent convection,
which increases with the luminosity and decreases with mass and effective temperature of the star~\citep{2011A&A...529L...8K}. This equation remains valid for red giant stars, as found in observations for which $V_{\rm n,max}$  varies from 10 to 800 cm s$^{-1}$~\citep{2013EPJWC..4303008S}. 
\begin{figure*}
\centering 
\includegraphics[scale=0.60]{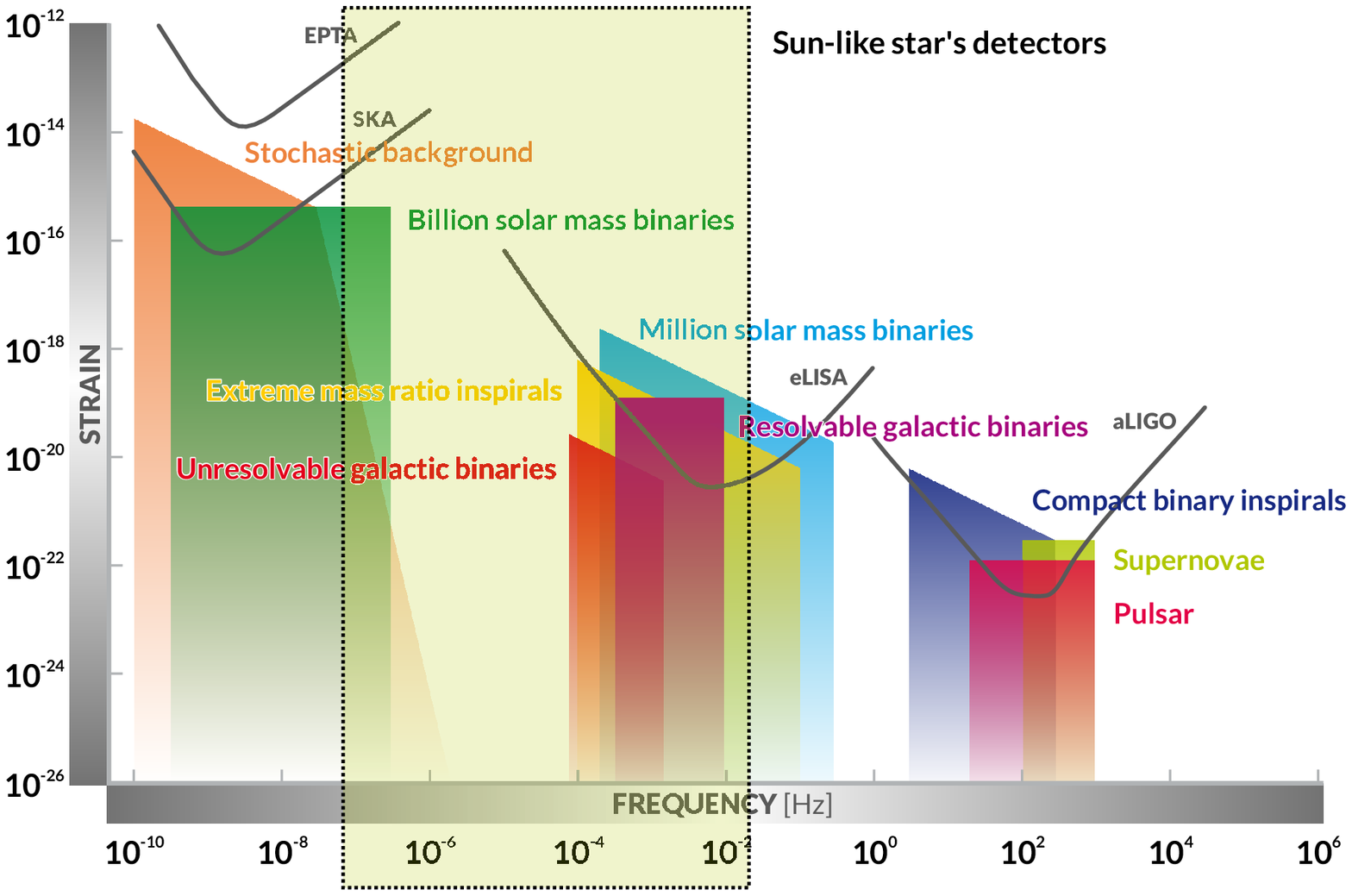}
\vspace{-2.0cm}
\caption{Comparison of {\it Sun-like star's detectors} with current gravitational waves detectors:
(i) The range of frequencies for probing for gravitational wave sources is evaluated from the observed 
acoustic spectra of  Sun-like stars: main sequence and subgiant stars show oscillations from  2 10$^{-4}$ Hz to 10$^{-2}$ Hz~\citep{2008ApJ...687.1180A};  and red giant stars from $10^{-7}$ Hz to $5\;10^{-3}$ Hz~\citep{2013sf2a.conf...25M}
(magenta shadowed area). (ii) The range of strain amplitudes possible to probe by {\it natural star detectors},  
depends mainly from the amplitude generated by the gravitational wave source, 
as unlike for gravitational wave detectors on Earth, Sun-like stars can be located quite close to the source. 
As for the example shown in Figure~\ref{fig:A} if a {\it natural star detector}
is located at a distance of 1 AU  from the gravitational wave source (red vertical line in Figure~\ref{fig:A}),  
for which $h_\star(1 AU)\sim 3\,10^{-14}$, this corresponds to $h_{\earth}\sim 2\,10^{-22}$ on Earth, 
leading to an effective strain gain of $10^{+8}$. Equally, if the {\it star detector} is located at
 1 solar radius from the gravitational wave source  $h_\star(1 R_\odot)\sim 5\,10^{-12}$.
At these distances all the low order modes ($n\le 4$) of the star are stimulated by the gravitational waves generated by the compact stellar binary. The strain sensitivity curve of the different 
gravitational detectors and most common sources of gravitational waves were adapted from~\citet{Moore:wk}.
}
\label{fig:B}
\end{figure*}

\smallskip

The acoustic mode frequencies of Sun-like stars  can probe gravitational waves in the range from 10$^{-7}$ Hz to 10$^{-2}$ Hz, which overlaps with the gravitational radiation frequency range that will be probed by eLISA~\citep{2013GWN.....6....4A} for the high frequency range and  EPTA~\citep{2010CQGra..27h4014F} and SKA~\citep{2007PASA...24..174J} in the low frequency range. 
More significant even is the possibility of probing the range from 10$^{-6}$ Hz to 10$^{-5}$ Hz, a frequency range for which no experiment has yet been planned. This is illustrated in Figure~\ref{fig:B}. In the low frequency range, this region  corresponds to the predictions of stochastic background radiation and supermassive binaries. In the high frequency range, it corresponds to unresolvable galactic binaries, extreme mass ratio inspirals and resolved galactic binaries~\citep{2009LRR....12....2S}.   
One of the targets of eLISA will be nearby ultracompact binaries, such as  the binary system AM CVn discussed in this article
as a template. The strategy of using Sun-like stars as detectors will enable us to
determine the impact of gravitational waves in the photospheric velocities of quadrupole modes,
or, at least, to fix an upper limit on the strain of the gravitational waves generated by these sources.
The $h$ sensitivity of eLISA will be $10^{-16}$ to $10^{-20}$ for the frequency range of 
$10^{-5}$ Hz to $10^{-2}$ Hz which is not sufficient to detect the gravitational waves produced by  the 
AM CVn binary system, predicted to have a strain amplitude on Earth  of $10^{-22}$  (cf. Figure~\ref{fig:B}).
Yet,  a future seismic mission could detect such gravitational waves in
stars like the Sun located at a distance of either 10 AU or 1000 AU 
from this binary system, if the $h_\earth$ is  of the order of $10^{-22}$ or $10^{-20}$, respectively.
As shown in figure~\ref{fig:A} the quadrupole modes of lowest orders will be stimulated by these
gravitational waves producing photospheric velocities possibly above the observational limit of the detector.
Moreover, other stars like sub-giant and red giant stars could scan other parts of the frequency range
of the gravitational wave spectrum, 
including outside of the current range of detectors.
Red giant stars with oscillations within the frequency range from  $10^{-7}$ Hz to $10^{-3}$ Hz~\citep{2013sf2a.conf...25M} can be used to explore events related with super-massive
binaries for which the strain on Earth is predicted to be of the order of  $10^{-14}$~\citep[see ][ and references therein]{Moore:wk}.

\section{Stellar systems and the detectability of gravitational waves}

Central to the applicability of using sun-like stars as detectors is to find an oscillating star  for which the photospheric velocity is above the threshold of detectability (cf. Figure~\ref{fig:A}).
As discussed previously for the sun's case, the photospheric velocity is of the order $10^{-8}\; {\rm cm\; s^{-1}}$ for a strain of $h_{-20}=1$ (see Table~\ref{tab:nusun}). Therefore for such a signal to be detected it is necessary to find a way to increase the photospheric velocity by 4 orders of magnitude. 
Two obvious possibilities are to choose stars for which their $\chi_n$ sensitive to gravitational waves is larger than the sun (like red-giants), or stronger gravitational wave sources (like the coalescence of black holes with a typical strain $\sim 10^{-17}$).  Here, nevertheless the discussion is focused in another point: the possibility of the next generation of stellar missions discovering gravitationally bound multi-body systems with a compact binary and an oscillating sun-like star, for which the seismic instrument has the necessary accuracy to measure the impact of gravitational waves in the acoustic oscillations.  
    
\smallskip
The probability of finding an oscillating star nearly enough to a compact binary is quite small, nevertheless, recently many gravitationally bound multi-body systems of three or more stars have been discovered, which gives some hope that one of such unique systems could actually be found.
The current theory  of stellar formation, predicts the existence of many  binary, triple  
and higher multiplicity stellar systems. This has also been confirmed by many astronomical 
observations. 
\citet{2014AJ....147...87T} studied the multiplicity of multi-body systems 
within 67 pc of the Sun and found that 13\% of stellar systems have three or more components. 
\citet{2010ApJS..190....1R} estimated that as much as 8\% are three-body systems, and
\citet{2013ApJ...768...33R} using {\it Kepler} data estimated that 20\% of close 
binaries have tertiary companions. Moreover,~\citet{2002ApJ...571..512H} have reported that within the first 20pc of the Sun,
25\% of white dwarf are in binary systems. 

\smallskip
Several triple systems have been discovered with characteristics relatively close
to the ideal case discussed in this article, compact binaries~\citep{2000A&A...358..417M} 
with a star companion. In particular, the COROT and {\it Kepler} missions 
have discovered several specific triple star systems~\citep{2014arXiv1410.8320S},  
like the KOI-126, a very small hierarchical triple system composed 
by a short-period binary of two 0.2$M_\odot$ stars orbiting by an evolved G-star 
(with a mass of 1.3$M_\odot$ and radius of 2.0 $R_\odot$)~\citep{2011Sci...331..562C}. 
The semi-major axis of the inner and outer binaries are 0.021 AU and 0.24 AU respectively.
Another equally different example is the triple system J0337+1715, a compact system  
with a total dimension smaller than 2 AU, where a close binary of a white dwarf and neutron star is 
orbited by a second white dwarf~\citep{2014Natur.505..520R}.  
Equally,~\citet{2014MNRAS.444L...1K} have discovered a close binary of white dwarfs
WD 0931+444 that possibly have an M dwarf companion, although for the moment sun-like oscillations
have not yet been discovered in M dwarfs~\citep{2015MNRAS.446.2613R}.  

Finally there is the possibility, that among the many binary systems formed by a white dwarf and sun-like star
discovered by the Kepler mission~\citep{2015arXiv150202303R,2014arXiv1410.8320S},   
like KOI-3278~\citep{2014Sci...344..275K}, the white dwarf could actually be a close compact binary. 
Equally, higher order multiplicity system could also be worth investigating to look for compact binaries
and oscillating stars~\citep{2015ApJ...799....4R}, like some 2+2 quadruple systems which 
have an outer separation of 500 AU. 

\smallskip
Although asteroseismology is not as precise as helioseismology,  if the optimal triple or higher order multi-body stellar system is found the accuracy of the future (and possibly present) stellar seismic  missions  
could be sufficient to detect gravitational waves.  Currently, the Plato mission observations for quiet stars,
expects that for an exposure time of only 15 minutes to be sufficient to average out the perturbing signal well below 
$1\; {\rm m\; s^{-1}}$, and in some cases the noise even gets down $10\; {\rm cm\; s^{-1}}$ in about 20-30 minutes~\citep{2013EGUGA..15.2581R}. It is possible that in cases of much longer time exposures, the compute photospheric velocities will be well below $1\; {\rm cm\; s^{-1}}$~\citep{2014ExA....38..249R}.    

Altogether, by using this proposed method, main-sequence, sub-giant and red giant stars increase 
significantly the detection power and the spectral range for identifying gravitational wave 
imprints in acoustic oscillations.

\section{Summary and Conclusion}
 
Stars like the Sun have the potential to become ideal detectors of gravitational waves.
Not only do they have a high scattering cross-section for gravitational waves, high quality factors and very likely  large photospheric velocities, but they also may be  located near strong gravitational wave candidates. 
Helio and asteroseismology allow us to probe these natural detectors with very high precision.
Indeed, for each oscillating star, the radial modes and dipole modes can be used to determine the oscillator characteristics like the damping rates, which is  critical to allow us to disentangle the stimulation caused by gravitational radiation from the excitation produced by turbulent convection.
The standard theory of stellar pulsations by turbulent convection establishes that low order acoustic modes ($l\le 3$) are equally excited and their excitation depends uniquely on the frequency of the mode and not on the degree. This model has already been proven to be consistent with observations for a large range of frequencies. Therefore, the radial and dipole modes with frequencies near the frequencies of the quadruple modes of low order 
can be used as reference to isolate the stimulation of 
low order quadrupole modes by gravitational waves from the stellar self-excitation.   

\smallskip
Although main-sequence, subgiant and red giant stars, all have the possibility to be good detectors of gravitational waves, subgiants and red giants have a substantial advantage in relation to main-sequence stars, as their rich spectrum has shown the existence of quadrupole mixed modes~\citep{2013ApJ...767..158B}.
These modes behave like gravity waves in the star's core and like acoustic waves in the stellar surface. Accordingly,  their eigenfunctions have large amplitude in both the core and the surface. As a consequence the modal length,  $L_n$ is much larger for these stars than in the Sun's case, so that their photospheric velocities are larger than the ones predicted for the Sun. 

The Sun and similar stars also have quadrupole gravity modes that can be excited by gravitational waves.
These modes have very large modal lengths $L_n$ compared with acoustic modes, because eigenfunctions of gravity modes have larger amplitudes in the star's core. Unfortunately, gravity modes are strongly attenuated when they propagate through the convection region and as such  it is very difficult to observe them in the star's surface. Actually, in the Sun
it has been a matter of dispute  whether a few gravity modes candidates were observed successfully~\citep{2004ApJ...604..455T}.  

\smallskip
The group of all oscillating stars in the solar neighborhood within one thousand parsecs radius,
constitutes the largest detector ever for gravitational radiation.   
Stars have some advantages over the current Earth detectors. There are thousands of oscillating stars scattered throughout space, some of which can be found relatively near gravitational wave sources. Alignments of  stars between the source and the Solar System can monitor the progression of gravitational waves throughout space, which can be use as a test to probe General Relativity, a goal that is difficult to achieve with present man-made detectors.

\smallskip
Equally important is the fact that such a new method to prove gravitational wave radiation could probe 
an important part of the spectral gravitational radiation window -- from one micro Hz to 100 micro Hz (cf. Figure~\ref{fig:B}), 
which is not probed by current  detectors and it is not expected to be probed by any planned future ones.
Actually, this is part of the reason why there are not many theoretical predictions for  gravitational 
radiation emitted in this spectral frequency range.  Among the possible candidates to produce the gravitational waves  
in this frequency range, there is the merger of intermediate mass black holes in the final year prior to the coalescence 
(for example a merger of a hundred solar mass black hole with an one million solar mass black hole) that
could originate from the high mass tail of the initial mass function of Population III stars~\citep{2001ApJ...551L..27M}.
The strain of these events varies from $10^{-23}$ to $10^{-17}$ depending on the chirp mass of the binary system in the coalescence phase~\citep{2004cbhg.sympE...4B}.

\begin{acknowledgments}
The authors would like to thank the anonymous referee for his/her inspiring questions 
and attentive reading of the manuscript. Moreover, the authors also acknowledge   
the comments and suggestions of Barry McKernan and Zoltan Haiman, 
the first authors that pointed out to us the importance of slow varying frequency  
gravitational waves on the excitation of stellar quadrupole modes.         
The work of I.L. was supported by grants from "Funda\c c\~ao para a Ci\^encia e Tecnologia"  
and "Funda\c c\~ao Calouste Gulbenkian". The research of J.S. has been supported at IAP by  
ERC project  267117 (DARK) hosted by Universit\'e Pierre et Marie Curie - Paris 6  
and at JHU by NSF grant OIA-1124403. We are grateful to the authors of ADIPLS and CESAM 
codes for having made  their codes publicly available. 
This research has made use of NASA's Astrophysics Data System.
\end{acknowledgments}
%
\bibliographystyle{yahapj}
\input{artILJS_biblo}


\end{document}

%% file: table_nusun_article.tex
\begin{table*} 
\centering
\caption{Quadrupole acoustic modes \\ Observational data and  Standard solar model}
\begin{tabular}{lllllllllllll}
\hline
\hline
n
&  Freq.~\footnote{The observational frequency corresponds to $\omega_n/2\pi$. The
table is obtained from  a compilation made by~\citet{2012RAA....12.1107T}, 
after the observations of~\citet{2000ApJ...537L.143B,2001SoPh..200..361G,2004ApJ...604..455T,2009ApJS..184..288J}.
The frequencies in italic correspond to theoretical predictions for the current standard solar model as in reference ~\citet{2013ApJ...765...14L}.} & 
$\eta_n$~\footnote{The damping rates are interpolated from a $\eta$ observational table 
of averaged values obtained for all global modes with $l\le 3$~\citep{1997MNRAS.288..623C}.
These observational results are consistent with the $\eta$ values of dipole modes measured by~\citet{2005AaA...433..349B}. 
The theoretical values of damping rates are  from~\citet{1999AaA...351..582H,2005AaA...434.1055G,2013ASPC..479...61B}.}  
 &   $\chi_{n}$ &    $L_n$ & $M_N$ & $\Sigma_n$   & $Q_n$  & $V_{n}$~\footnote{The photospheric velocity
is computed for a strain of $h_{\earth}=h_{-20}\;10^{-20}$ with $h_{-20}\sim 1$. }  & 
$T_n^{\star}$~\footnote{This value is computed for the compact binary system AM CVn (see Figure~\ref{fig:A} for details).} & $V_{n,u}$~\footnote{This photospheric velocity corresponds to the unsaturated limit
$T^\star_n \ll 1$ (or $\tau_f \ll \tau_n $). The values within ($\cdot$) are indicated for reference only.}  \\
 &  ($\mu Hz$) &  ($\mu Hz$)    &    ({\rm no-dim})  & $({\rm cm })$ & $(M_\odot)$ & (${\rm cm^{-2} \; Hz}$) & ({\rm no-dim}) & $({\rm cm \; s^{-1}})$ & ({\rm no-dim})  & $({\rm cm \; s^{-1}})$ \\
\hline
&                &   &  $\times 10^{-4}$& $\times 10^{7}$&  $\times 10^{-3}$ &  & $\times 10^{+8}$ & $\times h_{-20} $ 
&  $\times 10^{-3}$ & $\times h_{-20} $  \\
$f$   & $\mathit{347.10}$ & $\mathit{2.9\;10^{-7}}$  &  $-6.7432$ & $2.347$  & $0.5854$ & $94.931$ & $38$ & 
$1.4\,10^{-6}$ & $2.7$ & $7.1\;10^{-8}$ \\
$p_1$ & $\mathit{382.26}$ & $\mathit{2.9\;10^{-6}}$ & $-11.038$ & $3.841$  & $1.1148$  & $98.315$ & $4.1$ & 
$2.6\,10^{-7}$  & $23$  & $4.0\;10^{-8}$ \\
$p_2$ & $\mathit{514.48}$ & $\mathit{1.6\;10^{-5}}$ & $+2.1193$ & $0.737$  &  $0.1594$ & $0.9671$ & $1.0$ &
$1.7\,10^{-8}$  & $72$ &  $4.6\;10^{-9}$ \\
$p_3$ &  $\mathit{664.06}$ & $\mathit{7.4\;10^{-5}}$ &$-0.6286$ & $0.219$ &  $0.0466$ & $0.0424$ & $0.28$ &
$1.8\,10^{-9}$	 & $210$ & $8.2\;10^{-10}$ \\
$p_4$ & $\mathit{811.33}$ & $\mathit{2.6\;10^{-4}}$ & $+0.2133$ & $0.074$ &$0.0154$ & $0.0025$ & $0.10$ & 
$2.6\,10^{-10}$ & $506$ & $1.9\;10^{-10}$  \\
\hline 
&                &    &  $\times 10^{-6}$& $\times 10^{5}$   &   $\times 10^{-6}$ & &$\times 10^{6}$ &  &  $\times 10^{0}$   \\
$p_5$ & $\mathit{959.23}$ & $\mathit{7.9\;10^{-4}}$ & $-8.2377$ & $2.867$ & $1.0733$ &$2\;10^{-4}$ & $3.8$ &
$4.7\,10^{-11}$  & $1.1$ & ($5.0\;10^{-11}$)  \\
$p_6$ & $\mathit{1104.28}$& $\mathit{2.1\;10^{-3}}$ & $+3.4804$&$1.211$ & $0.4537$ & $2\;10^{-5}$ & $1.7$ & 
$9.8\,10^{-12}$  & $2.3$ & ($1.5\;10^{-11}$) \\
$p_7$ &$\mathit{1249.78}$ & $\mathit{5.3\;10^{-3}}$ &$-1.5051$ & $0.524$ & $0.1968$ &$2\;10^{-6}$ & $0.74$ &
$2.1\,10^{-12}$  &$4.7$ & ($4.6\;10^{-12}$)  \\
$p_8$& $1394.68\pm 0.01$ & $0.01$ & $+0.6836$& $0.238$& $0.0880$ & $2\;10^{-7}$  & $0.33$ &
$4.9\,10^{-13}$ & $9.4$ & ($1.5\;10^{-12}$)  \\
$p_9$& $1535.865\pm 0.006$ & $0.04$& $-0.3109$ & $0.108$ &  $0.0392$ &  $3\;10^{-8}$ & $0.12$& 
$8.9\,10^{-14}$  & $24$& ($4.4\;10^{-13}$)  \\
\hline 
&          &   &  $\times 10^{-8}$  &   $\times 10^4$ &  $\times 10^{-9}$ & &$\times 10^{4}$  & &  $\times 10^{1}$ 
&   \\ 
$p_{10}$ & $1674.534\pm 0.013$ & $0.08$   &  $+14.946$  &   $1.082$ &   $18.647$&    $3\;10^{-9}$ & $6.7$& $2.6\,10^{-14}$ 
& $4.1$ & $----$ \\
$p_{11}$ & $1810.349\pm 0.015$ & $0.10$   &  $-7.8242$   &   $0.520$ &   $10.06$&    $6\;10^{-10}$  & $4.1$&$9.0\,10^{-15}$ 
& $6.2$ & $----$ \\
$p_{12}$ & $1945.800\pm 0.02$  & $0.14$   &  $+4.3862 $    &   $0.272$ &  $6.106$&  $2\;10^{-10}$ & $2.9$& $3.8\,10^{-15}$  
& $8.3$ & $----$ \\
$p_{13}$ & $2082.150\pm 0.02$  & $0.21$   &  $-2.5981$  &  $0.153$&   $4.0176$&    $3\;10^{-11}$ & $2.1$& $1.7\,10^{-15}$  
& $10.9$ & $----$ \\
$p_{14}$ & $2217.69\pm 0.03$   & $0.31$   &  $+1.5564$   & $0.054$&   $2.7101$&    $9\;10^{-12}$  & $1.9$& $1.0\,10^{-15}$ 
& $11.4$& $----$ \\
$p_{15}$ & $2352.29\pm 0.03$   & $0.40$   &  $-0.9562$     &   $0.033$&  $1.9063$&  $3\;10^{-12}$  & $1.8$ & $6.4\,10^{-16}$
& $11.1$& $----$ \\
$p_{16}$ & $2485.86\pm 0.03$   & $0.44$   &  $+0.6204$   &   $0.022$& $1.4536$&    $1\;10^{-12}$ & $1.7$  & $4.2\,10^{-16}$
& $11.1$ & $----$\\
$p_{17}$ & $2619.64\pm 0.04$   & $0.53$   &  $-0.4180$   &   $0.014$& $1.1950$&    $4\;10^{-13}$  & $1.5$ &$2.6\,10^{-16}$ 
& $12.2$ & $----$ \\
$p_{18}$ & $2754.39\pm 0.04$   & $0.57$   &  $+0.2908$   &   $0.010$& $1.0383$&   $2\;10^{-13}$  & $1.5$ & $1.8\,10^{-16}$ 
& $12.0$ & $----$ \\ 
\hline
\hline
\end{tabular}

\label{tab:nusun}
\end{table*}   